\documentclass[aps,preprint,,amsmath,amssymb,superscriptaddress,longbibliography,floatfix]{revtex4-2}
\usepackage{graphicx}
\usepackage{dcolumn}
\usepackage{bm}
\usepackage{hyperref}
\usepackage[mathlines]{lineno}
\usepackage{algorithm}
\usepackage[noend]{algpseudocode}
\usepackage{xcolor}
\usepackage{pifont}
\usepackage{tikz}
\usepackage[normalem]{ulem}
\usepackage{filecontents}
\usetikzlibrary{matrix}

\usepackage{float}
\usepackage{soul}

\usepackage[final]{changes}
\definechangesauthor[name=GK, color=red]{GK}
\definechangesauthor[name=AG, color=red]{AG}
\usepackage{ulem}

\newcommand{\Torb}{T_{\mathrm{orb}}}
\newcommand{\Tspin}{T_{\mathrm{sp}}}
\newcommand{\TFL}{T_{\mathrm{FL}}}
\newcommand{\Ueff}{U_{\mathrm{eff}}}
\newcommand{\SRO}{Sr$_2$RuO$_4$\,}
\newcommand{\FTS}{FeTe$_{1-x}$Se$_x$\,}
\newcommand{\BFA}{BaFe$_2$As$_2$\,}

\begin{document}
\title{
The Hund-metal path to strong electronic correlations
}

\author{Antoine Georges}
\affiliation{Coll\`ege de France, PSL University, 11 place Marcelin Berthelot, 75005 Paris, France}
\affiliation{Center for Computational Quantum Physics, Flatiron Institute, 162 Fifth Avenue, New York, NY 10010, USA}
\affiliation{CPHT, CNRS, Ecole Polytechnique, Institut Polytechnique de Paris, Route de Saclay, 91128 Palaiseau, France}
\affiliation{DQMP, Universit{\'e} de Gen{\`e}ve, 24 quai Ernest Ansermet, CH-1211 Gen{\`e}ve, Suisse}

\author{Gabriel Kotliar}
\affiliation{Department of Physics and Astronomy, Rutgers University, Piscataway, New Jersey 08854, USA}
\affiliation{Condensed Matter Physics and Materials Science Department, Brookhaven National Laboratory, Upton, New York 11973, USA}

\begin{abstract}
Atomic physics has a profound impact on the physical properties of correlated electron materials. 
This article describes a prime example of this phenomenon. 
We provide a non-technical introduction to the physics of Hund metals, a broad class of materials 
which include in particular iron pnictides and chalcogenides, as well as oxides of the 4d transition-metal series such as ruthenates. 
We highlight experiments which reveal distinctive signatures of Hund physics in selected materials. 
A key property of Hund metals is a clear separation between the energy and temperature scales associated with 
spin and orbital degrees of freedom. 
%
We emphasize the conceptual and practical importance of the non-perturbative renormalization group flow which identifies the relevant degrees of freedom at each energy scale. 
The flow begins with local atomic degrees of freedom at high energy, displays a universal behavior 
in the intermediate regime of spin-orbital separation and ends into a broad diversity of possible ordered phases. 
Ordering can take place either as an instability of the Fermi liquid regime with coherent quasiparticles or 
directly from the intermediate regime before Fermi liquid quasiparticles had a chance to emerge. 
Dynamical mean-field theory provides a natural conceptual framework as well as a powerful computational method to explain, 
calculate and predict many physical properties of correlated materials such as Hund metals.\\
A copy-edited version of this article has been published (April, 2024) in Physics Today, 77(4), 46-53 (2024) 
\url{https://doi.org/10.1063/pt.wqrz.qpjx}

\end{abstract}


\maketitle

\newpage

\subsection*{Introduction}

Electrons in  solids  can behave either independently from one another or as collective
team players, resulting in emerging cooperative phenomena such as magnetism,
metal-insulator transitions or unconventional superconductivity,  to name a few. 
In the latter case, we talk about strongly correlated materials. 
\replaced[id=AG]{These materials raise}{which have posed}    
fundamental science  questions  while offering  promising  technological applications,  
\replaced[id=AG]{and have given}{giving}   
birth to a whole field of condensed matter physics. 
 
\replaced{In weakly correlated materials such as simple metals and semiconductors, electrons reside in very extended orbitals}
{The properties of weakly correlated materials such as simple metals and semiconductors are well described in a band picture with electrons in  very  extended orbitals} \replaced{and have a large kinetic energy}{and have a large kinetic energy}. 
We think of their quantum state in terms of  independent waves delocalized through the solid \added{(Bloch waves)}
\added{with an energy spectrum organized into bands}. 
In contrast, electrons in strongly correlated materials reside in more localized orbitals, hence  it is more natural to think of their quantum state 
in terms of correlated particles residing near the \replaced{nuclei}{atoms}. 
  
Strong correlation phenomena abound in materials  with partially filled d- and f-shells
such as  transition metals,  rare-earths,  and  actinides, but also in organic materials 
where electrons reside in molecular orbitals. The  degree of \replaced{correlation}{electron  localization} is \added{usually} controlled by pressure, \added{stress} or \deleted{by} doping.   
Recently, multilayers of two dimensional  
materials such as graphene or transition-metal dichalcogenides were turned into strongly
correlated  systems by twisting or misaligning the   layers  and applying gate voltages.  
 
Different families of materials follow distinct routes to strong correlation physics. In Mott systems, the motion of electrons is impeded and the kinetic energy  is blocked  by the  Coulomb repulsion. 
\deleted{An analogy is provided  by  a crowded  subway car during rush hours, where  it is difficult for passengers to move without stepping on the feet of a neighbor, a process that electrons with opposite spins perform for a very short time only, resulting in magnetic super-exchange.}     
Materials in the heavy fermion family have  two fluids of electrons which live rather independent  
lives at high temperatures: \replaced{mobile electrons }{itinerant spd conduction electrons} 
and localized f-electrons forming local magnetic moments. 
At very  low  temperature the hybridization or  quantum-mechanical  mixing between these two species of electrons becomes relevant. Then, a single fluid of itinerant \added{slowly moving `heavy'} electronic quasiparticles  emerges below a characteristic scale, the Kondo temperature. \deleted{These quasiparticles have an effective mass which can be as large as a thousand times  the bare electron mass}.  
 
\added{This article focuses on a new perspective describing} 
a broad family of materials whose properties cannot be understood within either the Mott or the heavy fermion paradigms.    
These materials display obvious signs of strong correlations but, unlike Mott systems they are very itinerant,  
\added{unlike simple metals they have local moments, and in contrast to heavy fermions they involve a single fluid of electrons}. 
\added{Hund  correlations are ubiquitous as they do not  require  tuning parameters  to the vicinity of  a phase transition.  
They are found in a  wide range of compounds with several energy bands crossing the Fermi level 
and involving  elements ranging  from  the transition-metal series to  the actinides.}
\deleted{They are multiorbital systems often comprising 3d and 4d transition metals and  have  multiple bands crossing the Fermi level. Unlike simple metals, they have local moments but, in contrast to heavy fermions, they involve a single fluid of electrons.}
As we shall see, a key character needed to understand these puzzling materials is the Hund’s rule coupling~\cite{Hund1925}. 

The  theory of what  we  now  call  `Hund metals', a term coined in Ref.~\cite{Yin2011}, was  launched by two pioneering papers that applied  Dynamical Mean Field Theory  to  the study  of the normal state of the iron based superconductors discovered in 2008~\cite{haule09} and to the celebrated 
three band Hubbard-Kanamori model~\cite{werner08}, a simplified  Hamiltonian that captures the essence  but not the specifics  of strongly  correlated systems.
It soon became apparent~\cite{demedici_janus_prl_2011, Yin2011} that a very large, and still growing, family of compounds fits within the new paradigm. Prominent members of this family are the ruthenium oxides.  
Their puzzling  \added{thermodynamic and transport} properties \added{and intricate phase diagrams}  were  studied since the mid-1990’s but a consistent physical picture of their normal state properties and the key role played by the Hund coupling only emerged recently~\cite{mravlje_Sr2RuO4_prl_2011}. 
The modern development of the Hund metal picture now explains the odd coexistence of itinerant and localized behavior noticed first \replaced{in photoemission studies} {on the example} of V$_5$S$_8$ by Fujimori and coworkers~\cite{Fujimori_V5S8}
and  has shed additional  light on classic experiments showing  
enormous variations  of the Kondo scale 
of  different magnetic atoms embedded in a  metallic host~\cite{Daybell_RevModPhys.40.380}.

In  this article we aim  to present in a pedagogical fashion the insights into Hund metals that have emerged over the last decade or so. We refer the reader to review articles~\cite{georges_Hund_review_annrev_2013}\cite{demedici_review,*demedici_capone_review}
\cite{stadler_review} for a more 
detailed exposition and appropriate references
to the large body of work and the vast community of scientists who have advanced this field. 

\begin{figure}[t!]
\begin{tabular}{c|c}
\includegraphics[width=0.45\textwidth]{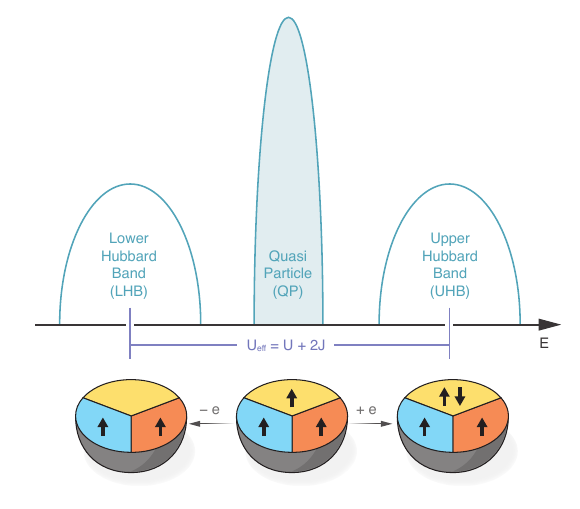}&
\includegraphics[width=0.45\textwidth]{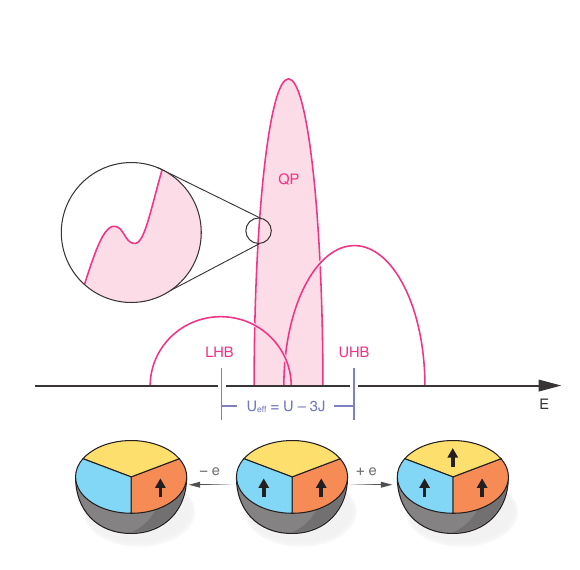}\\
\end{tabular}
\caption{
{\it 
Excitations in model Mott and Hund systems with 3 orbitals. 
The ground-state has 3 electrons (‘half-filled’ shell) in the left panel 
and 2 electrons in the right panel, each in a different orbital 
with parallel spins in accordance with Hund’s rule. Removing or adding an electron from the shell (bottom panels) leads to 
a lower Hubbard band (LHB) and upper Hubbard band (UHB) in the total density of states 
(plotted in the upper panel as a function of excitation energy E). 
The LHB and UHB are separated by an energy $\Ueff$. 
For a half-filled shell increasing the Hund coupling leads to
an increase of $\Ueff$ and of the Mott gap, while the opposite applies to non-half-filled shells, hence
promoting metallicity. For metals, the spectrum additionally contains low-energy
quasiparticle excitations (shaded peak) which can overlap with the Hubbard bands for non-half filled
shells when the Hund coupling is sizeable. 
For Hund metals, the QP peaks has a substructure due to spin-orbital separation, as depicted in the inset of the right panel.}
}
\label{fig:fig1}
\end{figure}

\subsection*{Atomic shells}

An atomic shell has  $M=2L+1$ orbitals corresponding to the possible values of the `magnetic' quantum number $L_z$.
Orbital degeneracy is lifted by the crystalline environment, leading to distinct subshells. 
Fig.~\ref{fig:fig1} illustrates the case of a $t_{2g}$ subshell with $M=3$ orbitals. 
Panel (a)  {describes  a  model for a metal in the Mott regime, with $N=3$ electrons in the 
ground-state (`half-filled' shell).}   
Panel (b) describes a model Hund metal  with ${N}=2$ electrons,
a situation relevant to Sr$_2$MoO$_4$ 
and also (exchanging electrons and holes) Sr$_2$RuO$_4$.
\deleted{An atomic shell has  $M=2L+1$ orbitals corresponding to the possible values of the `magnetic' quantum number $L_z$. 
Fig.~\ref{fig:fig1} illustrates this for $M=3$: in panel (a) the ground-state has $N=3$ electrons (`half-filled' shell) 
while it has $M=2$ electrons in panel (b). 
\deleted{The former is relevant e.g. to SrMnO$_3$, while the latter applies} to Sr$_2$MoO$_4$ 
and also (exchanging electrons and holes) Sr$_2$RuO$_4$. These materials have, respectively, three, two and four 
electrons in the $t_{2g}$ subshell of the Mo $4d$ orbitals with $L_{\mathrm{eff}}=1$. 
SrMnO$_3$ is a Mott insulator, while Sr$_2$MoO$_4$
and Sr$_2$RuO$_4$~\cite{mravlje_Sr2RuO4_prl_2011} are both 
Hund metals (with additional physics for the latter due to the proximity of a van Hove singularity). }
The electronic configuration of an atomic shell is determined by two key energy scales: 
the Hubbard energy $U$ and the Hund coupling $J$. 
When two electrons with opposite spins occupy the same orbital 
they feel the largest repulsive interaction $U$: the 
effective Coulomb interaction screened by the solid-state environment. 
In contrast, if two electrons occupy different orbitals, their effective repulsion is smaller: 
it is equal to $U^\prime<U$ if they have opposite spin, and to $U^\prime-J$ when they have parallel spin. 
$J$ is the intra-atomic exchange interaction, also called the `Hund's rule coupling'. 
When spherical symmetry approximately applies $U^\prime\simeq U-2J$, so that the interactions 
are in decreasing order: $U$, $U-2J$ and $U-3J$. 
\replaced{Given the magnitude of these interaction terms ($ J>0 $ ) }{Hence}, the lowest-energy configuration of $N$ electrons in an atomic shell with  $M$ orbitals is such that, 
if $N$ is smaller than $M$ (less than half-filled shell) each electron occupies a different orbital, all with parallel spins. 
If $N=M$ (`half-filled shell'), all orbitals are singly occupied, with maximal spin $S=N/2$ and  
if $N$ is larger than $M$, some orbitals must be doubly occupied. 
This is an illustration of more general rules formulated by Friedrich Hund in 1925~\cite{Hund1925}. 
A cartoon picture of Hund's rule is to think of a bus with rows of two seats each: if possible, 
passengers will typically spread themselves such as to have an empty seat next to them. 

\replaced {The  Hubbard  U   works like an   inverse  charge capacitance   penalizing   charge imbalance.    It is the main player  in  Mott  and heavy fermion materials.   It   blocks charge motion,   promotes localization in Mott systems, and reduces the  hybridization in heavy fermion materials.}
{$U$ is the main player in both Mott insulators and heavy fermion compounds: 
in the former it blocks charge degrees of freedom so that electrons remain localized on the atoms, while in the latter it reduces the effective hybridisation between 
the narrow and broad electron bands.} 
In contrast, in Hund metals, the Hubbard repulsion is too small to localize electrons. 
The key player is $J$ which forces the electrons to keep a collective configuration 
with the largest possible spin as they hop around, hence inducing strong correlations. 
In a nutshell, Mott insulators, heavy fermion compounds and Hund metals \replaced{are all correlated materials but their correlations arise 
from charge, hybridisation and spin blocking, respectively}{Mott insulators, heavy fermion compounds and Hund metals are associated with charge
blocking, hybridisation blocking and spin blocking, respectively.
}.

\replaced{In  a solid, an  atomic configuration with one extra electron added to the
ground-state  can  move around.  This  excitation is  known as the ‘upper Hubbard band’ 
}{In the solid, the electronic configuration of each atom changes as the electron hops from one atom to the next. 
%
An excited atomic configuration with one extra electron added to the ground-state can move, 
forming a broad band known as the `upper Hubbard band' }(UHB).  
Similarly, excitations with one less electron form the `lower Hubbard band' (LHB) - see Fig.~\ref{fig:fig1}. 
These bands can be viewed as the result of atomic transitions $d^N\rightarrow d^{N\pm 1}$ broadened by the solid-state environment. 
The electron removal (respectively, addition) can be measured by photoemission (respectively, 
inverse photoemission) spectroscopy.  

The gap between the LHB and the UHB can be estimated from the energy cost $\Ueff=E_0(N+1)+E_0(N-1)-2E_0(N)$ 
for transferring a single electron from one atom to another.   
One finds $\Ueff=U+2J$ for a half-filled shell: 
in that case increasing the Hund coupling leads to an {\it increase} of the Mott gap. 
This explains why many materials with a  half-filled shell are Mott insulators. 
In contrast, for a non half-filled shell $\Ueff= U^\prime-J\simeq U-3J$ and the Mott gap {\it decreases} upon increasing $J$. 
As a result, 
$J$ promotes metallicity by increasing the overlap between Hubbard bands.
These effects were pointed out early on by van der Marel and Sawatzky~\cite{vandermarel_sawatzky_prb_1988} 
and Fujimori~\cite{Fujimori_V5S8}  
- see \cite{georges_Hund_review_annrev_2013}\cite{demedici_review,*demedici_capone_review} 
for reviews and additional references. 

\subsection*{Quasiparticles}

In metals, there are additional excitations at low-energy, 
\replaced{called Landau quasiparticles. 
These quantum states are waves delocalized through the solid, with a well-defined momentum. 
The occupied and the unoccupied states are separated by a surface in momentum space called the Fermi surface. 
The group velocity of the quasiparticle states near this Fermi surface can be strongly reduced by interactions, 
corresponding to an enhancement of the `effective mass' of these excitations. 
}
{which propagate like coherent waves.  They are long-lived and have a well-defined Fermi surface. 
Their velocity can be strongly affected by interactions - think of a passenger trying to move inside a crowded subway car. 
Comparing the observed velocity to that of the same system in the absence of correlation effects defines 
the interaction-induced `effective mass enhancement' $m^*/m$.} 
%
\deleted{These excitations are called quasi-particles because they resemble the infinite-lived Bloch waves 
of non-interacting systems, but have finite lifetimes and  renormalized properties.
The Landau theory of Fermi liquids postulates that all low-energy excited states of the many-body system 
can be constructed by combining together quasiparticles, like elementary bricks of  a Lego game.} 

\replaced{The quasiparticle weight $Z$ is the  probability of an electron to be in a quasiparticle state 
when removed (or added) to the ground-state of the solid, as probed in photoemission (inverse photoemission). 
In a weakly correlated material, most of the excitations are quasiparticles, and $Z$ is close to unity (see Fig.2).  
In contrast, a small value of $Z$ is a smoking gun for strong correlations.}
{The quasiparticle weight $Z$ 
is the fraction of the total addition and removal spectrum corresponding to quasiparticles: 
a small value of $Z$ is also a smoking gun for strong correlations.}  
In metals close to the Mott transition, quasiparticles emerge between Hubbard bands. 
In contrast, in Hund metals, the quasiparticle excitations emerge from overlapping 
Hubbard bands (see Fig.~\ref{fig:fig1}),   
making it harder to distinguish between the two types of excitations in spectroscopic experiments. 

\subsection*{Dynamical mean field theory}

\replaced{
In weakly correlated materials, most of the excited states can be expressed in terms of Landau  quasiparticles. 
This is not the case in correlated electron materials: hence a different strategy is required. 
A proper description starts from the many-body eigenstates of individual atoms (atomic multiplets), 
involving energy scales of several electron-volts, which constitute a sizeable part of the excitation spectra.  
}
{
Hence, in order to understand strongly correlated materials, it is crucial to start from a proper description 
of the many-body eigenstates of individual atoms (atomic multiplets), involving energy scales of several electron-volts. 
Proceeding towards lower energy scales, the theory should describe if and how quasiparticle excitations emerge. 
This typically happens on much lower scales: for Sr$_2$RuO$_4$ a coherent Landau Fermi liquid 
is formed only below $T_{\mathrm{FL}}\sim 25$~K, corresponding to an energy scale of about 20 milli-electronvolts. 
}
This perspective is a serious revision of the standard `condensed matter textbook': 
instead of an electron gas with interactions, we 
should think of a material as a collection of atomic quantum many-body systems which exchange electrons between them.

This point of view is at the heart of {\it dynamical mean-field theory} 
(DMFT, see \cite{georges_review_dmft,*kotliar_dmft_physicstoday} for reviews). 
DMFT focuses on \replaced{the sequence of quantum jumps between}{the history of local changes in the} electronic 
configurations of the atoms as electrons hop between them. 
\added{It describes this process as the emission and absorption of electrons between the 
atom and an effective self-consistent bath representing the rest of the system
}

\deleted{A major success of} The theory describes 
\replaced{how quasiparticles emerge at low-energy as a result of this process.}
{the system evolves as one follows the flow from high-energy particle like atomic excitations 
to low energy wave-like quasiparticle excitations.}
The quantum mechanical duality between high-energy particle-like atomic excitations 
and low-energy wave-like quasiparticle excitations is at the heart of the physics of strongly correlated materials. 
\deleted{As detailed below, the flow from high energy to low energy is particularly rich and complex for Hund metals, 
and DMFT proved crucial in elucidating it.} 
%
\replaced{Combined with electronic structure methods,}{Thanks to the combination of many-body and electronic structure computational methods, }
DMFT \deleted{also} provides a practical framework to understand and predict the properties of quantum materials  
starting from their structure and chemical composition. 
\added{It is now part of a broader class of `quantum embedding' methods which treat 
different degrees of freedom in the solid at different levels of quantum mechanical accuracy.
}

\begin{figure}[t!]
\includegraphics[width=0.9\textwidth]{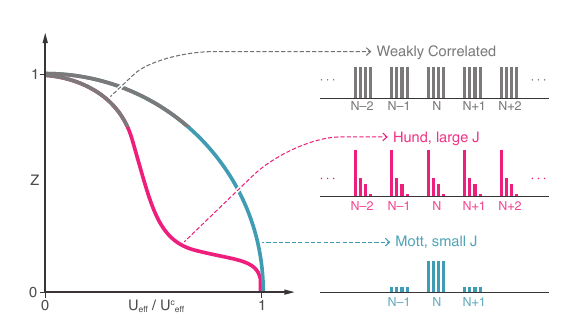}
\caption{
{\it 
Schematic dependence of the quasiparticle weight $Z$ on the ratio $\Ueff/\Ueff^c$, with $\Ueff^c$ 
the critical coupling where the Mott transition takes place. The plot compares three cases: a weakly correlated metal (grey), a strongly correlated metal close to the Mott transition (blue) and a metal with substantial Hund coupling (pink). 
On the right hand side of the figure we see the valence histograms which represent the probability 
of different atomic configurations in the ground-state by the height of the vertical lines.  
Each group of vertical lines corresponds to a given number of electrons $N$, ($N, N\pm 1, N\pm 2, \cdots$). 
In Hund metals, 
the configurations with maximal spin have a higher probability (as indicated by the 
prominent spike in each charge sector).
}
}
\label{fig:fig2}
\end{figure}

\subsection*{Hund metals}

Figure~\ref{fig:fig2} illustrates three different metallic regimes: a weakly correlated metal with $Z$ close to unity and 
two strongly correlated metals with a small value of $Z$, one in the Mott regime and one in the Hund regime. 
When the Hund coupling is sizeable and the shell is neither singly occupied nor half-filled, 
the dependence of $Z$ on $U_{\mathrm{eff}}$ displays a plateau.  Hence, there is 
an extended regime displaying strong correlations while the system is not close to the Mott metal-insulator transition. 
Quoting 
Ref.~\cite{demedici_janus_prl_2011}, the Hund coupling is `Janus-faced': 
on the one hand it pushes the Mott transition away while on the other hand it leads to an extended strongly 
correlated metallic regime by decreasing $Z$ and enhancing \replaced{the quasiparticle effective mass}{\,$m^*/m$}. 

A useful diagnostics of these different regimes which can be performed 
within DMFT is the histogram 
of the relative weights associated with each atomic configuration in the ground-state of a given material. 
This is what an observer embedded  in the solid at a given atomic site would record by measuring 
the time that an atom spends  in each atomic configuration.  
As depicted on Fig.~\ref{fig:fig2}, this histogram  extends over many different charge configurations and atomic multiplets 
for a weakly correlated metal. 
In contrast, in the Mott regime, it is concentrated on a single charge state $N$ with small contributions from $N\pm 1$. 
For a Hund metal, the histogram extends over many charge states but it is dominated within each charge sector 
by the multiplets with large spin and angular momentum consistent with Hund rules, hence providing a 
direct signature of `Hundness'~\cite{haule09}.

\subsection*{Go with the flow}

\begin{figure}[t!]
\includegraphics[width=0.9\textwidth]{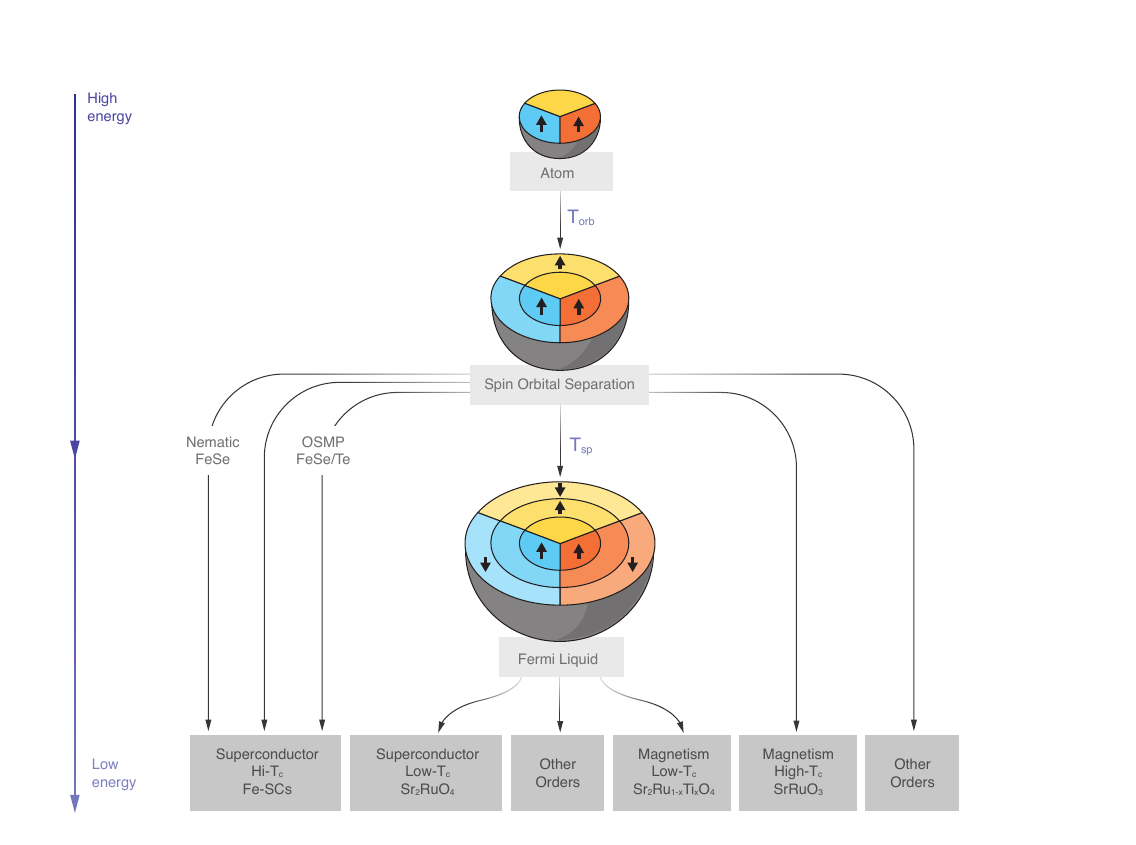}
\caption{
{\it
Schematic flow from the high-energy regime where atoms fluctuate between different multiplet configurations down 
to low-energy where coherent quasiparticle excitations emerge.  
In Hund metals, the quenching of orbital and spin fluctuations occurs at distinct 
temperatures, $\Torb$ and $\Tspin$.  
The DMFT description of this process is depicted here schematically as an `onion-like' sequence of 
effective energy shells added to the isolated atom. 
At low-energy in the Fermi liquid regime 
all orbitals and spin states are effectively filled by an equal number of electrons, so that 
all atomic degrees of freedom are quenched. 
In the spin-orbital separated (SOS) regime between $\Torb$ and $\Tspin$, 
orbitals are quenched but a higher spin is transiently formed ($S=3/2$ in this example). 
The figure also illustrates the large diversity of symmetry breaking instabilities and long-range orders that emerge towards the end of the flow.  
Importantly, these instabilities can either take place at low temperature 
within the coherent Fermi liquid regime as in Sr$_2$RuO$_4$ or at higher temperatures 
from the Hund metal regime when the  coherent Fermi liquid is not formed yet 
as for example in  SrRuO$_3$. 
}}
\label{fig:fig3}
\end{figure}

Fig.~\ref{fig:fig3} illustrates the evolution of the active degrees of freedom as a function of energy scale 
(or, using renormalisation group parlance, the `flow'), 
from a high-energy regime with fluctuating spin and orbitals associated with atomic multiplets,  
down to low energy where these degrees of freedom reorganize into itinerant quasiparticles. 
DMFT describes this process as the gradual binding of bath electrons to atomic orbital  
and spin degrees of freedom. 
Powerful renormalisation group analyses 
by two groups, - a collaboration between  Ludwig Maximilian University in Munich,  Brookhaven National Laboratories and Rutgers University, and a team from the Jo\v{z}ef Stefan Institute in Ljubljana 
have recently provided a theoretical understanding and numerical computation 
of this flow\cite{aron_2015,horvat_2016}. 
For a review and additional  references, see \cite{stadler_review}.

As illustrated on Fig.~\ref{fig:fig3}, this flow involves two distinct crossover temperatures: 
a higher one $\Torb$ at which the local orbital degrees of freedom 
\replaced{become itinerant}{are quenched} 
and a lower one $\Tspin=\TFL$ at which the spin degrees of freedom also 
become coherent and delocalized, 
corresponding to the formation of the Fermi liquid. 
This spin-orbital separation (SOS) can lead to a distinctive 
feature in one-particle spectra~\cite{stadler_review}, see Fig.~\ref{fig:fig1}.

\begin{figure}
\includegraphics[width=0.8\textwidth]{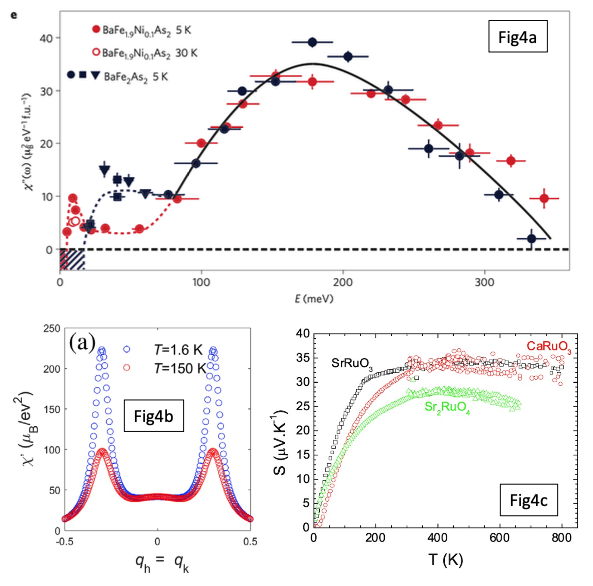}
\caption{
{\it 
Experimental signatures of Hund metals from INS (inelastic neutron scattering) and transport measurements.  
(a): Energy dependence of $\int d\mathbf{q}\, \chi^{\prime\prime}(\mathbf{q},\omega)$ for pristine \BFA  at $T=5$~K and   
5\% Ni-substituted \BFA at $5$~K and $30$~K~\cite{liu_2012}.
(b): Momentum dependence of the susceptibility $\chi^\prime(\mathbf{q},\omega=0)$ of \SRO~ 
extracted  from a fit to  INS data~, at temperatures $T=1.5$~K and $150$~K 
\cite{steffens_2019,*Strand_2019} .
(c): Seebeck coefficient (thermopower) vs. temperature for SrRuO$_3$, CaRuO$_3$ and \SRO
~\cite{klein_mrs_2007,*mravlje_seebeck_sr2ruo4_2016}. 
Notice the quasi universal value of the plateau, which corresponds to fluctuating spins 
but quenched orbital 
fluctuations, a distinctive signature of Hund metals discussed in the text. 
}
}
\label{fig:fig4}
\end{figure}

\deleted[id=AG]{The broad crossover region between $\Torb$ and $\Tspin$ is the Hund metal regime.} 
\added[id=AG]{
In the SOS regime which is a hallmark of Hund metals}, 
the spins are quasi-localized, leading to a Curie-Weiss temperature dependence $\chi \sim1/(T+T_0)$ of 
the uniform magnetic susceptibility. 
In contrast, the orbital and charge degrees of freedom are itinerant and the corresponding 
susceptibilities are temperature-independent (Pauli-like). 

\subsection*{Smoking guns}

Inelastic neutron scattering (INS) is an important experimental tool  in this context, as it  probes directly  the nature of the spin excitations.
The frequency dependence of the imaginary part of the  spin susceptibility, integrated over  wavevector, is related to the density of states of spin excitations. 
Multiplying it by  a Bose factor, and integrating it over frequency  gives the size of the fluctuating local moment. 
It is  displayed in Fig.~\ref{fig:fig4}(a) 
for several members of the iron based superconductor 122 family~\cite{liu_2012}. 
The high energy part of the susceptibility and the fluctuating local moment,  does not vary much from one material to another or as temperature is changed. 
%
In contrast 
\replaced{the low-energy regime}{This regime} displays a strong temperature and material dependence: 
pristine \BFA undergoes a transition to a stripe phase while the nickel-doped compound becomes superconducting. 
This illustrates the large diversity of \deleted{instabilities and} possible long-range orders 
at the low-energy end of the flow of Fig.~\ref{fig:fig3} \added{which} 
\deleted{These ordering} 
can emerge in two ways. 
Either at $T<\TFL$ as a low-temperature instability \added{of the Fermi liquid} resulting from interactions between coherent quasiparticles, 
or at a higher temperature above $\TFL$ \replaced{as an instability of}{within} the less coherent Hund metal.  
This latter case is hard to treat with textbook many body methods and requires a framework such as DMFT for its description. 

\SRO is an example of the former: it undergoes a low-temperature 
superconducting instability at $T_c\sim 1.4$~K, well within the 
Fermi liquid regime which is attained at $\TFL \sim 25$~K, 
as evidenced by the quadratic temperature 
dependence of the resistivity as well as quantum oscillations.  
Above $\TFL$, this material enters the SOS regime. 
The ferromagnet SrRuO$_3$ (Curie temperature $T_c\sim 160$~K) 
and the nematic superconductor FeSe (with nematic $T_c\sim 80$~K and  superconducting
$T_c\sim 9$~K) are examples of the latter.

At intermediate temperatures or energies, 
both \SRO and SrRuO$_3$ as well as other members of the ruthenate family 
display characteristic features of Hund metals.
Recent resonant inelastic X-ray scattering (RIXS) measurements performed by Hakuto Suzuki and coworkers in the group 
of Bernhard Keimer at the Max Planck Institute of Stuttgart, Germany, provide direct experimental evidence of the 
separation of energy scales associated with orbitals and spins in \SRO \cite{Suzuki-24}. 
Evidence that spins are fluctuating while orbitals are not in the SOS regime  
was previously provided by the measurement of the thermoelectric Seebeck coefficient. 
The latter measures the electrical potential drop when a sample is subject to a thermal gradient, and  
its value can be related to the entropy of fluctuating degrees of freedom in the material. 
%
%
The Seebeck coefficient of three different ruthenates is displayed on Fig.~~\ref{fig:fig4}(c). 
It increases linearly with $T$ at low temperature as expected in a 
Fermi liquid metal, but it reaches a plateau at higher temperature, with a value 
$\sim 30\,\mu\mathrm{V}\cdot\mathrm{K}^{-1}$  which depends weakly on the compound. 
As noted by Y.~Klein, S.~H\'ebert and coworkers \added{from the CRISMAT laboratory in Caen, France} 
and confirmed by DMFT calculations~\cite{klein_mrs_2007,*mravlje_seebeck_sr2ruo4_2016} 
this value can be explained by considering that spin degrees of freedom are quite localized and that their entropy provides the only 
contribution to the Seebeck coefficient. Including also orbital degrees of freedom 
would lead to a negative value~\cite{mravlje_seebeck_sr2ruo4_2016}, in contradiction to experiments. 

\added{INS experiments also confirm the presence of fluctuating local moments in \SRO.  
Indeed, the recent experiments} of P.~Steffens and coworkers\cite{steffens_2019,*Strand_2019} 
show that the momentum dependence 
of the \deleted{zero frequency} magnetic susceptibility 
of \SRO ,displayed on Fig.~\ref{fig:fig4}(b), can be decomposed into two components: 
one which is broad in momentum space 
and weakly temperature dependent and another one which is peaked at specific wave-vectors and increases 
upon cooling. 
\added[id=AG]{DMFT calculations have revealed that} 
the former can be associated with local moments and Hund physics   
while the latter arises from itinerant quasiparticle excitations\cite{steffens_2019,*Strand_2019}. 
The itinerant component grows and sharpens upon cooling, reflecting a tendency towards a spin-density wave instability. 
In the pristine material, this instability is arrested  (the correlation length reaching 
only a few lattice spacing at low-$T$), but small amounts of 
chemical substitutions or strain induce a spin density wave order at very low temperatures 
illustrating  again the low energy diversity of the flow of Fig.~\ref{fig:fig3}.
%
%


\begin{figure}[t!]
\includegraphics[width=\textwidth]{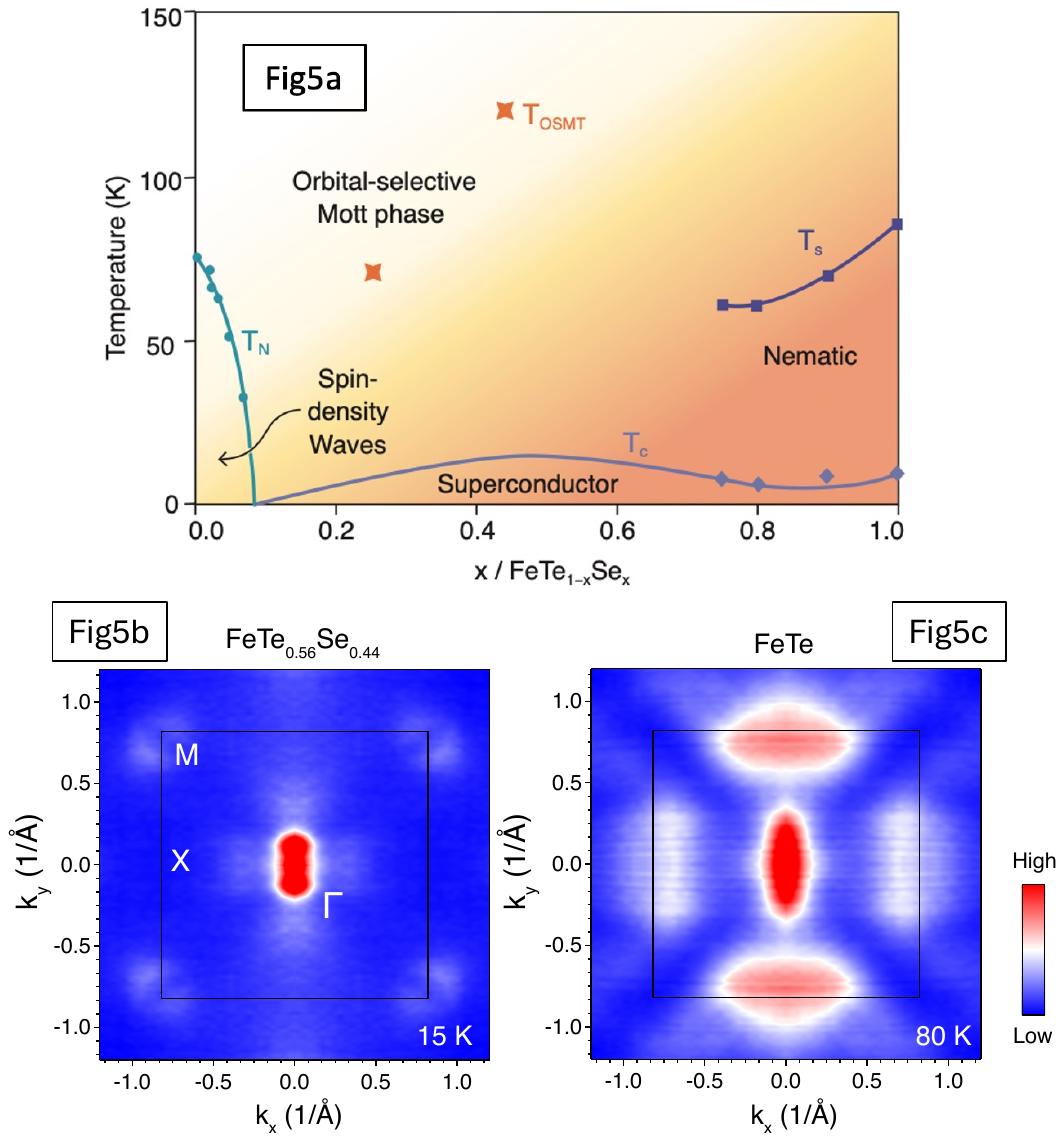}
\caption{
{\it 
\replaced{Orbital selective Mott phenomenon in \FTS.}
{Angular Resolved Photoemission Spectroscopy (ARPES) view of the orbital selective Mott phenomenon.}
(a): Phase diagram as a function of concentration $x$ and temperature. 
The two red data points mark the boundary of the orbital selective Mott phase (OSMP). 
A lighter color indicates a lesser degree of coherence of the quasiparticles. 
(b): Fermi surface (FS) of FeTe$_{0.56}$Se$_{0.44}$ at $15$~K in the  correlated metallic phase, 
as determined by Angular Resolved Photoemission Spectroscopy (ARPES). 
(c): FS of  FeTe at $80$~K (in the OSMP). 
Notice the new FS pockets around the X point in the OSMP, due to the de-hybridization of the $d_{xy}$ orbital as it undergoes the orbital selective Mott transition. 
Figure adapted from Ref.~\cite{Huang2022}, courtesy of Ming Yi and Jianwei Huang. 
}
}
\label{fig:fig5}
\end{figure}

\subsection*{Orbital differentiation}

As the concept of Hund metals emerged, it became clear that the Hund coupling plays a key role in promoting {\it orbital differentiation}, 
namely the emergence of a large difference in the degree of correlations among otherwise similar orbitals~\cite{demedici_review,*demedici_capone_review}.  
This phenomenon is especially relevant to iron-based superconductors~\cite{demedici_OD}. 
It has been observed with different experimental probes such as Scanning Electron Microscopy (STM) by the group 
of S\'{e}amus Davis and  ARPES (angle resolved photoemission) by the group of Z.~X.~Shen. 
We illustrate this phenomena in Fig.~\ref{fig:fig5} using experiments on \FTS from the group of Ming Yi~\cite{Huang2022}. 
As the tellurium-rich end of the phase diagram is approached, the \added{electrons residing in the} 
$d_{xy}$ orbital become much heavier and 
incoherent than for the others $d$-orbitals, as predicted by DMFT calculations\cite{Yin2011}. 
Eventually the system enters 
an  `orbital selective Mott phase' (OSMP) - an extreme form of orbital differentiation in which this orbital is so incoherent that its dispersion  cannot be observed while the remaining $d$ orbitals still form itinerant quasiparticle bands.
Fig.~\ref{fig:fig5} reveals a dramatic change of the Fermi surface as the OSMP 
regime is entered. In this regime, the hybridization of the $d_{xy}$ orbitals with other orbitals 
is turned off, resulting in changes in the quasiparticle dispersions:  
comparing panels (b) and (c) in Fig.~\ref{fig:fig5}, we see that 
new Fermi surface sheets emerge around the $X$ point of the Brillouin zone in the OSMP. 

\subsection*{Outlook}

Finally, we emphasize that the crucial role played by the Hund coupling in inducing strong electronic correlations 
is relevant to many more materials than the ones discussed in this article. 
Hund physics may actually occur in combination or in competition with other important factors  
such as the proximity of a van Hove singularity \added{(a region of momentum space with vanishing electronic velocity)}, 
as in \SRO \cite{georges_Hund_review_annrev_2013},  
or the emergence of Mott physics, for example when bringing Fe-based 
superconductors closer to a half-filled $d^5$ shell by hole doping
~\cite{demedici_review,*demedici_capone_review}.

In all cases, to describe  Hund metals it is essential to think in terms of the dynamical fluctuations between 
different many-body atomic configurations. Hund metals therefore provide 
one of the most beautiful illustrations of the DMFT concept which pictures a solid as self-consistently embedded atoms. 
DMFT provides a deep understanding of strongly correlated materials 
by describing the flow from 
\replaced{fluctuations between atomic configurations}{atomic fluctuations} 
at high energy to emerging coherent quasiparticles at low energy. 
A variety of instabilities and symmetry breaking into ordered phases 
typically emerge along this flow (Fig.~\ref{fig:fig3}). 

Along with this conceptual framework and in combination with electronic structure methods, DMFT also 
equips us with a powerful computational framework to calculate and predict thermodynamic, transport and 
especially spectroscopic properties of correlated materials. 
Indeed,  evidence for the crucial role of Hund physics largely emerged from 
an extraordinary number of  successful comparisons  between  experimental studies and materials specific calculations by  
groups across the world. 

A frontier in this area is to extend these methods both in scope and accuracy in order 
to reach lower temperatures and the lower energy range of the phase diagram. 
A particularly important and interesting  challenge is to understand how 
symmetry breaking instabilities such as high-temperature superconductivity emerge out of a metallic state 
\deleted[id=AG]{such as the Hund metal regime}  
in which a description in terms of long-lived coherent quasiparticle excitations is not applicable. 
\added{At low temperature, longer range spatial correlations must be taken into account and 
computational methods also face more severe algorithmic challenges.}
This will \added{therefore} require the development of new techniques which will both advance and extend 
the scope of quantum embedding methods to reach lower energies and address fluctuations over longer length scales.

 {\it Acknowledgements --} We are grateful to Luca de’ Medici, Olivier Gingras, Fabian Kugler, Jernej Mravlje, and André-Marie Tremblay for their suggestions and insightful comments, and to Pengcheng Dai, Sylvie Hébert, Jianwei Huang, Sinjie Xu, and Ming Yi for sharing their experimental data and helping to display them. We are also very grateful to Lucy Reading-Ikkanda for creating figures in the article. G.K. was supported by the National Science Foundation under grant  NSF DMR-1733071. The Flatiron Institute is a division of the Simons Foundation. 


%

\end{document}